# Learning-based multiplexed transmission of scattered twisted light through a kilometer-scale standard multimode fiber

*Yifan Liu [a)], Zhisen Zhang [a)], Panpan Yu, Yijing Wu, Ziqiang Wang, Yinmei Li, Wen Liu [b)], and Lei Gong [b)]*

Department of Optics and Optical Engineering, University of Science and Technology of China, Hefei 230026, China
[a)] Y. Liu and Z. Zhang contributed equally to this work.
[b)] Author to whom correspondence should be addressed. Email: Wenliu@ustc.edu.cn; leigong@ustc.edu.cn

Multiplexing multiple orbital angular momentum (OAM) modes of light has the potential to increase data capacity in optical communication. However, the distribution of such modes over long distances remains challenging. Free-space transmission is strongly influenced by atmospheric turbulence and light scattering, while the wave distortion induced by the mode dispersion in fibers disables OAM demultiplexing in fiber-optic communications. Here, a deep-learning-based approach is developed to recover the data from scattered OAM channels without measuring any phase information. Over a 1-km-long standard multimode fiber, our method is able to identify different OAM modes with an accuracy of more than 99.9% in the parallel demultiplexing of 24 scattered OAM channels. To demonstrate the transmission quality, color images are encoded in multiplexed twisted light and our method achieves decoding the transmitted data with an error rate of 0.13%. Our work shows the artificial intelligence algorithm could benefit the use of OAM multiplexing in commercial fiber networks and high-performance optical communication in turbulent environments.

## I. INTRODUCTION

Light has several degrees of freedom, *e.g.* wavelength[1], polarization[2], amplitude that can be used to encode data. Recently, the orbital angular momentum[3] (OAM) of light has been recognized as a unqiue degree of freedom to increase the data capactiy[4-6]. The OAM is carried by light beams with helical wavefront, termed vortex beams or twisted beams. Such beams with different OAM values are orthogonal to each other. This orthogonality enables multiple independent twisted beams to carry unlimited data channels. For communication purpose, these beams need to be multiplexed and transmitted over long distances in either free-space or fibers[7-12]. However, the free-space twisted light transmission is strongly influenced by dyanmic atmospheric turbulence and light scattering. Especially, the light scattering will scramble the wavefronts of the OAM modes and destroy the orthogonality between the OAM channels[13]. The fiber-based soluitons could overcome the influence of dynamc atmospheric turbulence and potentially support long-range transmisison. However, the mode coupling and dispersion in common fibers, which resembles the process of multiple scattering, distort the tranmitted spatial beams and prevent the use of OAM multiplexing in fiber-optic communications.

To apply OAM multiplexing, different types of fibers with a special ring-shape structure[14-17] are preferred for stable OAM transmission, which face greater manufacturing challenge and larger fiber loss compared to standard multi-mode fibers (MMFs). Lately, J. Wang et. al. proposed to use few-mode fiber couplers at the ends of a standard MMF to excite and select target OAM modes[10,18]. By using the specially-designed couplers, OAM multiplexed transmission through a standard MMF was achieved. Nonetheless, the couplers support only a few OAM modes, hindering the high-demensional OAM multiplexing. More recently, the wavefront shaping techniques that enables focusing beyond scattering are introduced for spatial mode transmission through standard MMFs. For instance, a vectorial time reversal technique realized transferring 210 Laguerre-Gauss modes over an 1-km-long fiber[19]. A transmission matrix-based light field imaging technique achieved OAM-multiplexed transmission through a 35-m MMF[20]. However, such techniques involved measurement and control of the full complex field transmitted through the fiber, which complicates the long-range communication links. Furthermore, machine learning provides another solution to addressing the distortion in fibers, but hitherto previous works mainly focused on optical imaging[21,22] or transfer of images through an MMF[23,24]. For optical transmission with OAM multiplexing, earlier demonstrations are limited to recognition of individual OAM modes undergoing weak atmospheric turbulence[25,26] or misalignment[27], but have not achieved recognition of scattered OAM modes. So far, direct OAM multiplexed transmission through a long standard MMF remains challenging.

In this work, a deep-learning-based approach is proposed to precisely extract encoded data from multiplexed OAM beams that transmitted through a long standard MMF. Propagation of twisted light along an MMF strongly scrambles its wavefront, yielding a random speckle pattern at the output. Our approach, termed OAM-Demux-Net, can learn the relationship between the output speckle pattern (intensity only) and the input OAM fields (amplitude and phase information) by using a convolutional neural network (CNN). As such, the network is able to undo the scrambling, discriminate each scattered OAM channel, and directly output the encoded data. To test its validity, a fiber-optic transmission system with a 1-km-long standard OM3 MMF is built based on a digital micromirror device (DMD) that allows us to encode multiplexed OAM channels. Trained with experimental data, our approach achieves to identify OAM channels with an accuracy of > 99.9% in the parallel demultiplexing of 24 scattered OAM channels. In addition, its performance is further demonstrated by transferring image data at a 24-bit rate through the fiber and a transmission error rate as low as 0.13% is achieved. Our work paves the way for direct use of mode-division multiplexing in existing fiber-optic networks and high-performance optical transmission in scattering environments.

## II. CONCEPT AND PRINCIPLE

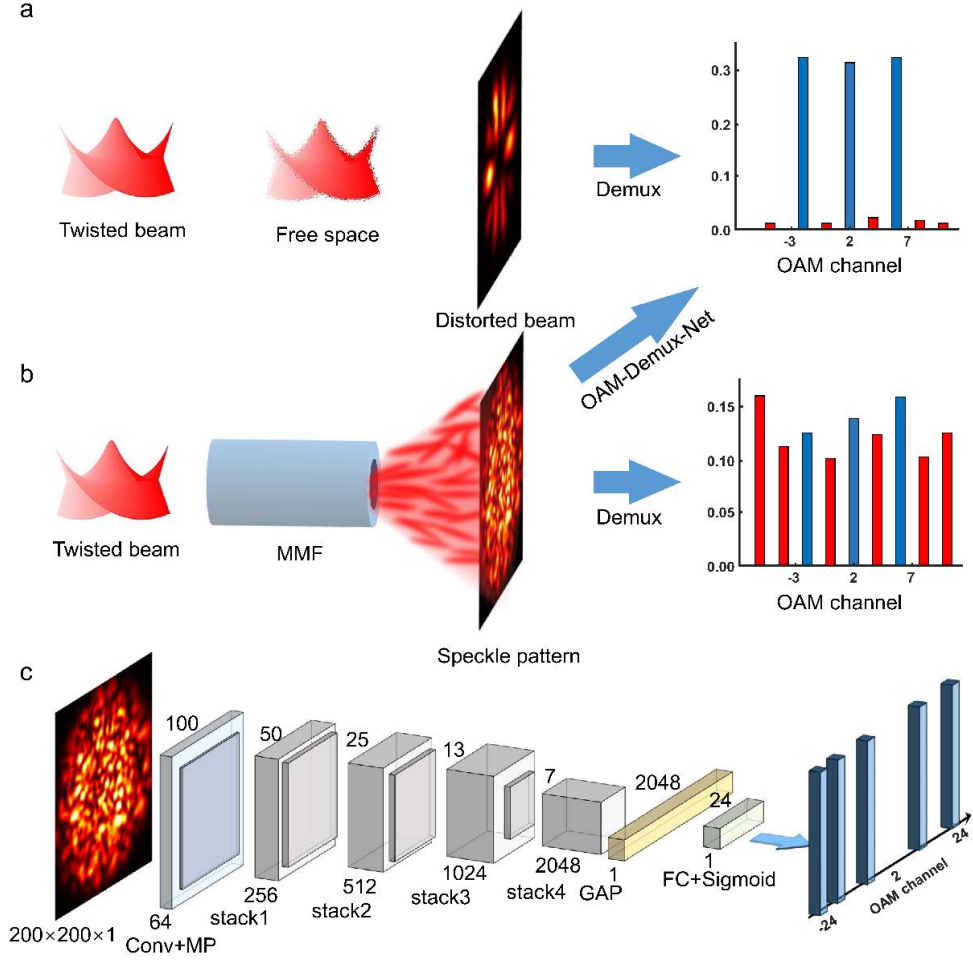

**FIG. 1.** Concept of demultiplexing scattered OAM channels with deep learning. (a) Free-space OAM communication link, where all the OAM channels can be identified at the receiver. (b) The MMF-based OAM multiplexed tranmission link. The data-carrying twisted light propagating along the MMF is scrambled by the mode dispersion, resulting in a random speckle pattern at the receiver. The scattering and scrambing disables the OAM demultiplexing with conventioanl techniques, whereas our OAM-Demux-Net allows precise discrimination of each OAM channel. (c) Schematic diagram of the OAM-Demux-Net. Conv: convolution layer; MP: max pooling layer; GAP: global average pooling layer; FC: fully connected layer.

The OAM-carrying beams possess a helical phase front $\exp(il\phi)$, and the OAM beams with different topological charges, $l$, are intrinsically orthogonal. Thus, multiple data-carrying channels can be created with different OAM beams. For OAM-multiplexed transmission, the total data channels scale with the number of OAM beams utilized for multiplexing. In practical communication links, each OAM beam carrying an independent data stream co-propagates with other OAM beams. To extract the transmitted data, all the OAM channels must be identified precisely at the receiver, as illustrated in Figure 1a. When undergoing strong scattering or intermodal coupling, these independent OAM channels are scrambled and mixed so that only the speckle pattern can be detected at the receiver. As a consequence, the channels cannot be recognized by conventional demultiplexing techniques (Figure 1b). To undo the mixing for the OAM identification, a CNN-based deep learning network is constructed and trained to recover the transmitted data by addressing all OAM channels from the speckle pattern.

Figure 1c shows the structure of our CNN model, termed OAM-Demux-Net. The network is based on the architecture of residual CNN[28,29], which allows us to efficiently extract the features of the object under study and reduce the quantity of parameters. In this model, the speckle intensity patterns are set as inputs, which are fed to the first convolution layer with a $7\times7$ kernels and (2, 2) strides, followed by a $3\times3$ max pooling layer with (2, 2) strides. After the convolution layer, the feature maps are sent into four stacks with 3, 4, 6 and 3 residual blocks, respectively. After the four residual blocks, a global average pooling layer and a fully connected layer with a sigmoid classifier are attached. Such a network enables high-accuracy recognition of scattered OAM channels after training. The OAM demultiplexing is actually a multi-class multi-label classification problem. In such a classification, our network directly outputs the labels of all the OAM channels used for multiplexing.

### III. EXPERIMENTAL SETUP

To validate our approach, we built a fiber-optic transmission system using a 1-km-long standard MMF (Corning ClearCurve, OM3 -125 μm), as illustrated in Figure 2. A He-Ne laser ($\lambda$=633 nm; Thorlabs, HNL210LB) is utilized as the light source, which is not shown in the figure. The output laser beam is expanded and collimated to fully illuminate the surface of the DMD (Vialux, V-9501) with an incident angle of 24 degrees. To generate arbitrary OAM beams, the DMD shapes the wavefront of the incident beam via a 4-$f$ configuration and a spatial filter in the Fourier plane. The shaped wavefront is dictated by the binary hologram displayed on the DMD, which is calculated by using the superpixel method[30,31] to encode both the amplitude and phase of light. The generated OAM beam is then coupled into the fiber via an objective lens (10X, NA = 0.25; Olympus). After propagation through the fiber, the transmitted beam was collected by another objective lens with the same NA and detected by a CMOS camera (D752, PixeLINK). The captured intensity images are set as the inputs of our OAM-Demux-Net.

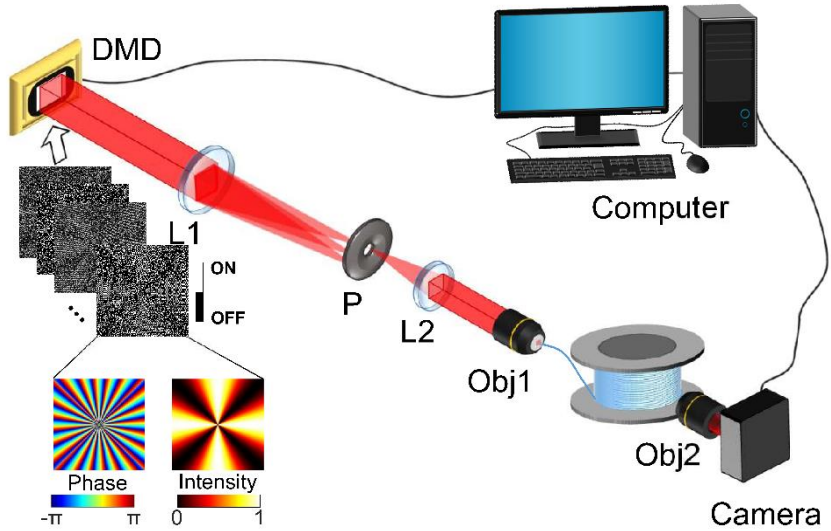

**FIG. 2.** Sketch of the fiber-optic OAM-multiplexed transmission system. L: lens; P: pinhole filter. Insert presents the binary holograms that are projected onto the DMD to shape the multiplexed OAM fields.

## IV. RESULTS AND DISSCUSSIONS

### A. Demultiplexing of scattered OAM channels with OAM-Demux-Net

For model training and testing, two groups of datasets are acquired in terms of 16 and 24 multiplexed OAM channles. The OAM bases with an interval of 1 (*e.g.*, $l = -8, -6, \cdots, 8$) is used for 16-channel multiplexing, and an interval of 2 (*e.g.*, $l = -24, -22, \cdots, 24$) is set for 24-channel multiplexing. For each group, different combinations of OAM bases are multiplexed and transmitted through the MMF, and each combination corresponds to an OAM superposition mode. Theoretically, there are in total $2^N$ superposition modes, where $N$ is the number of the OAM bases used for multiplexing. In practice, part of the combinations are randomly selected for the training and testing. Here, 30000 and 100000 combinations are used in the tranning for 16 and 24 multiplexed channels respectively, while 20000 ones that are never participated in the training process are employed as the testing data. In the experiments, the selected superposition modes are generated and switched by the DMD. A series of speckle images for all the input modes are collected at the receiver as the raw data. These images are downsampled to 200×200 to accelerate the processing procedure.

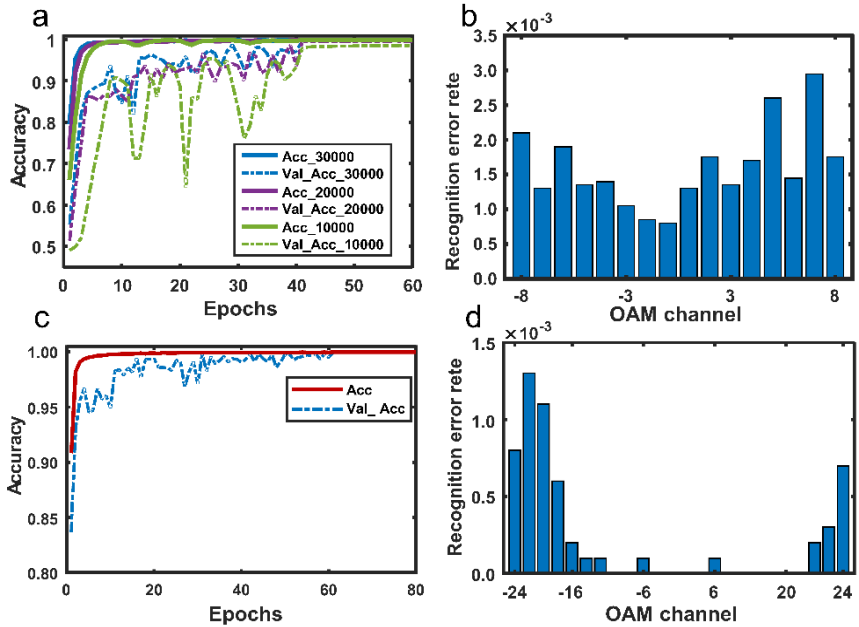

**FIG. 3.** Accuracy of parallel demultiplexing of multiple OAM modes with the OAM-Demux-Net. (a, c) The training accuracy curves of 16- and 24-mode demultiplexing. (c, d) The corresponding recognition error rate of each OAM mode using the testing dataset.

In the training process, the input dataset is divided into training and validation subsets according to a ratio of 8:2. It is a iterative process to learn the weight parameters, and 60 and 80 epochs are adopted for 16- and 24-OAM mode demultiplexing, respectively. Between two iterations, a loss function of binary crossentropy and an Adam optimizer are used. The initial learning rate is set to be 0.001, and is lowered by 10 times at epoch 40 and epoch 60. The accuracy for training is shown in Figure 3a,c. Obviously, the accuracy increases with the the number of iterations for the training dataset, and our network achieves a final validation accurcay of > 99.9%. To further analyze the learning abilty, we reduce the scale of the input dataset and perform the training. For example, the scale is reduced from 30000 to 20000 and 10000 for 16 OAM bases. The corresponding accuracy curves are also presented in Figure 3a. Accordingly, the validation accuracy becomes 99.6% and 98.6%, suggesting the good robustness of the network even with reduced learning dataset scale.

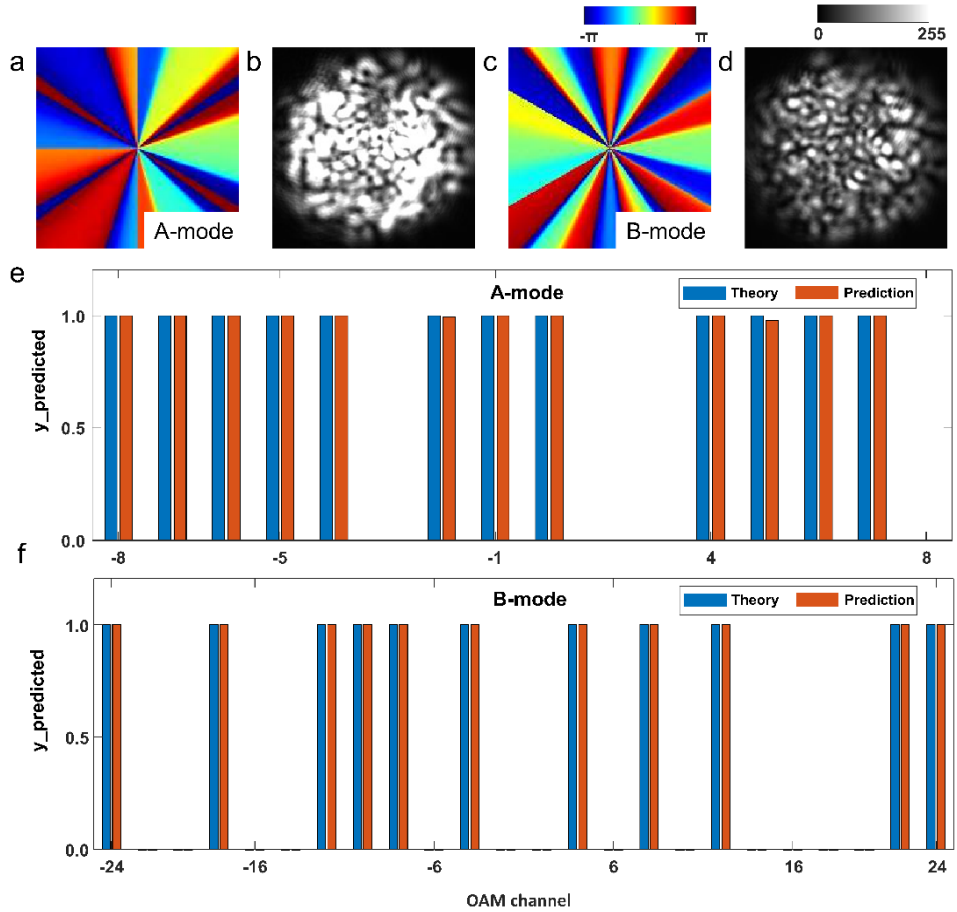

**FIG. 4.** The results of recognizing OAM bases from the speckle patterns. (a, c) The wavefront of two multiplexed OAM modes transmited through the MMF. (b, d) The corresponding spackle patterns received at the end of the fiber. (e, f) The demultiplexing results from the intensity speckle patterns using our trained network.

After training, the performance of our network is tested with the testing dataset in the parallel demultiplexing of scattered OAM channels. Figure 3b,c shows the recognition error rate of each OAM mode. The error rate of single channel in recognizing 16 multiplexed channels is lower than 0.29%, while it is lower than 0.13% in recognizing 24 channels. This implies that a large interval between two adjacent bases leads to a low recognition error. Besides, it is found that the accuracy of recognizing high-order OAM modes is a little bit lower than the low-order modes because propagation of the high-order modes are more sensitive to the mode dispersion in fibers. In spite of this, the trained network achieves an averaged recognition accuracy of close to 99.98% in the parallel demultiplexing of 24 scattered OAM channels. For further examination, we present the detected speckle patterns and the predicted results by our network for two tested OAM modes in Figure 4. The predicted results conincide well with the theoretical results. For recognizing the OAM channels, a binary threshold criterion is used in our network. The testing results verify that our approach successfully recognizes the multiplexed OAM modes transmitted through a km-scale standard MMF with an overall recognition rate of close to 100%.

**B. OAM-encoded data transmission through a 1-km-long standard MMF**

We further applied the trained network for OAM-multiplexed data transmission. As a demonstration, a color parrot image with a 220 × 220-pixel resolution was encoded pixel by pixel into multiplexed OAM modes and then transmitted through the fiber-optic system. Any color could be expressed by a weighted mixture of three primary colors (*i.e.*, red, green and blue) as shown in Figure 5a. For each color, a 256-level data can be encoded by using a binary digital byte with 8 bits, where every bit takes a value of 0 or 1. Thus, an OAM superposition state containing 24 OAM bases (*i.e.*, $l_n=\pm24, \pm22, …, \pm4$ and $\pm2$) enables to encode any RGB color pixel, where every eight bases correlate with one primary color. Then the pixel information can be transferred using a one-time transmission with this OAM superposition state. Figure 5b shows the results of the transmitted picture with an error bit rate of 0.13%, which is defined as the ratio between the incorrect number of pixels and the total number of the image.

For further examination, the error rates corresponding to different primary color channels are shown in Figure 5c. The error rate of green color channels is lower than that of red or blue color channels which are encoded with higher-order OAM bases. Nevertheless, the overall low error rate of the three color channels guarantees the high-fidelity transmission of image data through the 1-km-long standard MMF.

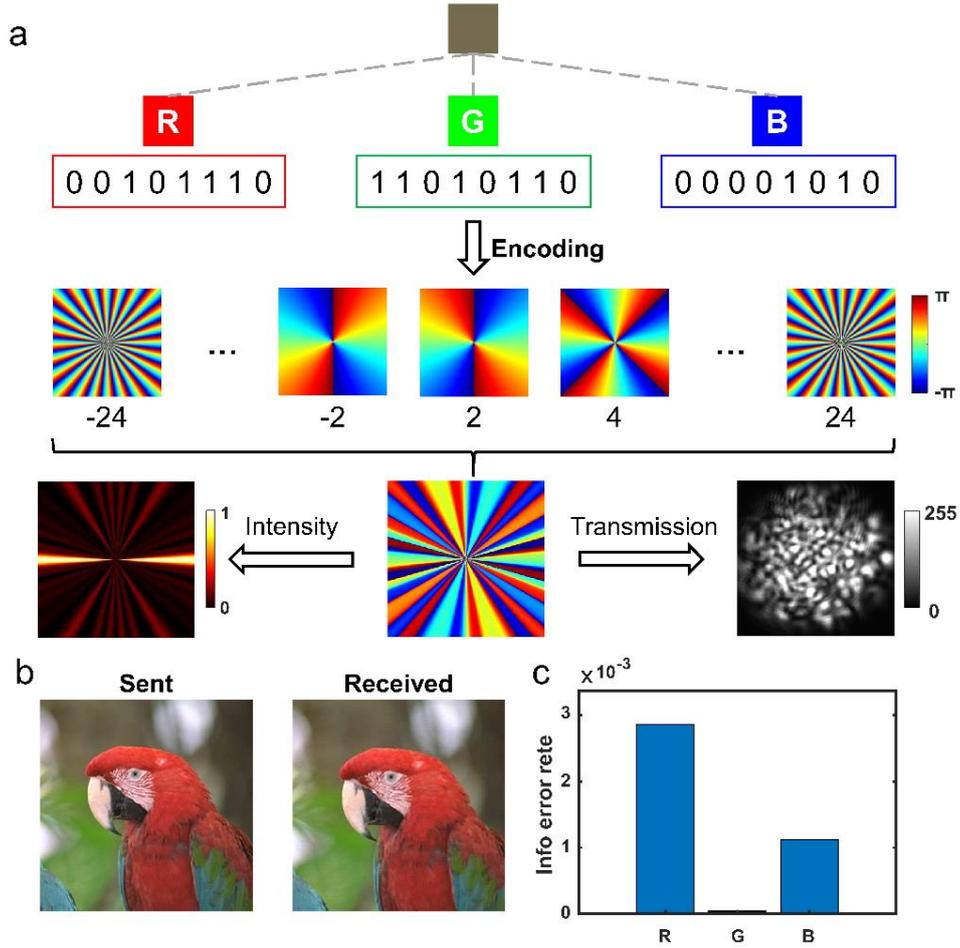

**FIG. 5.** Results of OAM-encoded data transfer through a standard 1-km-long MMF. (a) RGB encoding scheme using 24-OAM multiplexing. (b) Example of sent and received color images (Parrot, 220 × 220 pixels) in the data transfer experiment. (c) The information error rates in different RGB channels.

## V. CONCLUSION

In summary, we have proposed a CNN-based approach termed OAM-Demux-Net for demultiplexing the scattered OAM channels and applied for twisted lights transmission through a km-scale MMF. After training with experimental data, our approach achieves recognition of scattered OAM modes with an accuracy of >99.9% in the parallel demultiplexing of 24 OAM channels in a 1-km-long standard MMF. Furthermore, the trained network is utilized to transfer OAM-multiplexed data through a self-built fiber-optic link. In the experiment, color images are encoded and transferred via 24 multiplexed OAM channels and our network achieves the scattered data recovery with an error rate of 0.13%. Based on such network, multiplexing more spatial modes[25,32] or other degrees of freedom of light[33] (*e.g.*, wavelength, polarization) could be expected to realize high-bandwidth optical communication. Our work thus paves the way for applying mode multiplexing in existing fiber-optic networks.

**ACKNOWLEDGES**

This work was supported by National Natural Science Foundation of China (NSFC) under grants 11974333, 31870759, and 31970754 as well as the Hefei Municipal Natural Science Foundation (Grant No. 2021001). P.Y. thanks support from the China Postdoctoral Science Foundation (2021M703114).

**Conflict of Interest**

The authors declare no conflict of interest.

**Data Availability**

The data that support the findings of this study are available from the corresponding author upon reasonable request.